\documentclass[twocolumn,showpacs,preprintnumbers,amsmath,amssymb,superscriptaddress,prb]{revtex4-1}
\usepackage{graphicx}
\usepackage{dcolumn}
\usepackage{bm}

\begin{document}

\title{
Dependence of atomic arrangement on length of flat bands \\
in zigzag BC$_2$N nanoribbons
}

\author{Tomoaki Kaneko}
\email{KANEKO.Tomoaki@nims.go.jp}
\affiliation{Computational Materials Science Unit, NIMS, Sengen 1-2-1, 
Tsukuba 305-0047, Japan}

\author{Kikuo Harigaya}
\email{k.harigaya@aist.go.jp}
\affiliation{Nanosystem Research Institute, AIST, Higashi 1-1-1, 
Tsukuba 305-8565, Japan}

\date{\today}

\begin{abstract}
We theoretically study the electronic properties of BC$_2$N nanoribbons with zigzag edges using a tight binding model. 
We show that the zigzag BC$_2$N nanoribbons have the flat bands and edge states when atoms are arranged as B-C-N-C along the zigzag lines. 
The length of the flat bands in the wavevector space depends on the atomic arrangement.
This property can be explained by the deviation of the linear dispersion of the BC$_2$N sheet from K point of the honeycomb lattice.
The charge distributions in the edge states depend on the atomic arrangement.
We also show that the charge distribution of the edge states in zigzag BC$_2$N nanoribbons where the outermost sites are occupied with B and N atoms is different from those in conventional graphene zigzag edge.
Such charge distribution causes different magnetic structures.
We investigate the magnetic structure of BC$_2$N nanoribbons with zigzag edges using the Hubbard model within a mean field approximation.
At the zigzag edge where the outermost sites are occupied with B and N atoms, ferromagnetic structure appears when the site energies are larger than the on-site Coulomb interaction.
\end{abstract}

\pacs{73.20.At, 73.21.Cd, 73.22.-f}

\maketitle

\section{Introduction}

Graphene is an atomically thin carbon sheet in which carbon atoms are arranged in a honeycomb lattice.  Due to its outstanding electronic structure and electron transport properties, graphene attracts much interests for future electronic devices. 
Nanocarbon materials which are fractions of graphene exhibit different electronic properties compared with graphene. Graphene nanoribbons are finite width graphene sheets, which are the one of the famous examples of nano-carbon materials.\cite{Fujita1996jpsj,Nakada1996prb}
Recently, graphene nanoribbons have been fabricated by e-beam lithography\cite{Han2007prl}, unzipping of carbon nanotubes\cite{Jiao2009nature,Kosynkin2009nature} and using bottom-up processes \cite{Cai2010nature}.

The electronic properties of graphene nanoribbons strongly depend 
on the edge structures.\cite{Fujita1996jpsj,Nakada1996prb,Wakabayashi1999prb,Miyamoto1999prb,Son2006prl}
Graphene nanoribbons with armchair edges have the band 
gaps.\cite{Son2006prl}
Graphene nanoribbons with zigzag edges have the so-called flat bands at the Fermi level.\cite{Fujita1996jpsj,Nakada1996prb,Wakabayashi1999prb,Miyamoto1999prb}
The states corresponding the flat bands are localized at the zigzag edges.
They are the edge states\cite{Fujita1996jpsj,Nakada1996prb,Wakabayashi1999prb,Miyamoto1999prb}
In the honeycomb lattice, there are two inequivalent sites, A- and B-sublattices.
For the formation of edge states, these sublattice structures play decisive roles.  The distribution of electronic charge of the edge states becomes finite only one sublattice sites including the outermost sublattice.\cite{Fujita1996jpsj,Nakada1996prb}
Quite recently, the edge states in graphene nanoribbons have been confirmed by STM/STS measurement.\cite{Tao2011nphys}

On the other hand, hexagonal boron-nitride (BN) sheet also shows the honeycomb lattice structure in which B and N atoms are arranged in A- and B-sublattice sites alternately.
Due to their chemical difference between B and N, BN sheet has a wide band gap.\cite{Catellani1987prb,Blase1995prb}
Since B (N) atoms behave as acceptors (donors), boron-carbon-nitride--graphene sheet doped with B and N--should show interesting controllable electronic properties by doping. 
Ci {\it et al}.\ recently have reported the synthesis of hybridized BN and graphene sheets using thermal catalytic chemical vapor deposition. \cite{Ci2010nmat}
Such hybridized BN and graphene sheets show different electronic transport properties compared with graphene and BN sheets. \cite{Ci2010nmat,Song2012prb}
The electronic properties of BN and graphene hybridized nanoribbons have been investigated by several authors. \cite{Nakamura2005prb,He2010apl,Basheer2011njp,Kan2008jcp}
In such nanoribbons, ferrimagnetic ordering along the edges were predicted.\cite{Nakamura2005prb,He2010apl,Basheer2011njp,Kan2008jcp}
Recently, Kaneko {\it et al}.\ reported that the appearance of flat bands and edge states in zigzag BCN nanoribbons where the outermost C atoms of graphene nanoribbons are replaced with B and N atoms alternately.\cite{Kaneko2012arXiv-alternation}
They reported that the distributions of charge and spin densities in the edge states are different from those in the edge states of zigzag graphene nanoribbons, i.e., the electronic charge distributes over both sublattices and the spin density can become ferromagnetic at the edges.

BC$_2$N sheet is organic analogous of graphene, which can be regarded as one of example of boron-carbon-nitride.
Graphite-like BC$_2$N has been synthesized using chemical vapor depositions of boron trichloride, BCl$_3$, and acetronitrile, CH$_3$CN as the starting materials.\cite{Kouvetakis1989SynthMet,Sasaki1993ChemMater}
The electronic properties of BC$_2$N sheets depend on the atomic arrangement.\cite{Liu1989prb,Nozaki1996JPCSolids,Azevedo2005EurPhysJB}
The electronic properties of nanoribbons made with BC$_2$N have been also investigated by several authors.\cite{Lu2010jpcc,Lu2010apl,Xu2010prb,Lai2011Nanoscale}
Lu {\it et al}.\ reported that BC$_2$N nanoribbons can have the magnetization even for armchair edge nanoribbons.\cite{Lu2010jpcc,Lu2010apl}
Xu {\it et al}.\ showed that BC$_2$N nanoribbons with zigzag edges have the linear dispersion when the atoms are arranged C-B-N-C in the transverse direction.\cite{Xu2010prb}
However,there are no reports on the presence of the flat bands and edge states in BC$_2$N nanoribbons. 
The purpose of this paper is to explore the flat bands and edge states in BC$_2$N nanoribbons.

\begin{figure*}[t!]
\centering
\includegraphics[width=13cm]{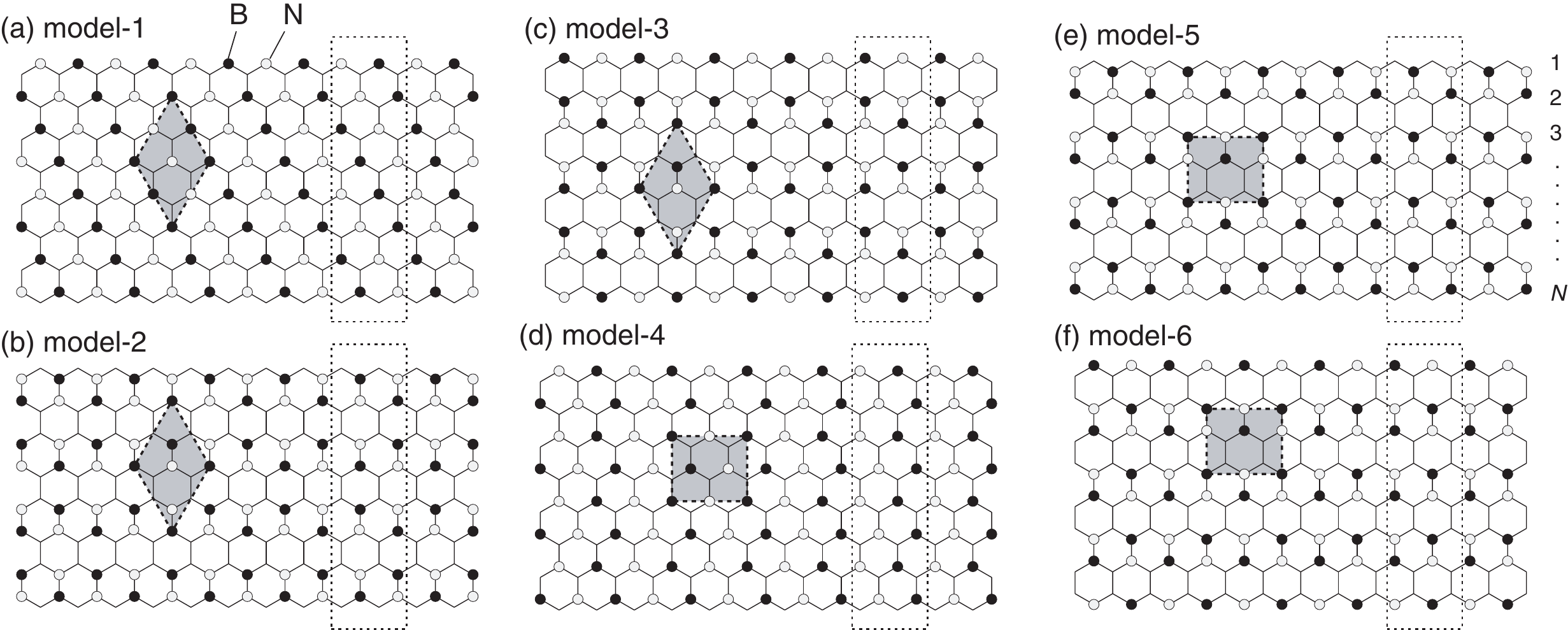}
\caption{
Models of BC$_2$N nanoribbons handled in this study.
In this figure, the black and white circles represent B and N atoms, respectively.
C atoms are located at the empty vertices.
The dotted rectangles represent the unit cells and the shaded regions represent the unit cell of BC$_2$N sheets.
}
\label{fg:RibbonStructure}
\end{figure*}

In this paper, we will investigate the electronic properties of BC$_2$N nanoribbons with zigzag edges using a tight binding model.
We will show that zigzag BC$_2$N nanoribbons have the flat bands and edge states when atoms are arranged as B-C-N-C along the zigzag lines.
The charge distribution of the edge states in zigzag BC$_2$N nanoribbons is different from those in conventional graphene zigzag edge.
Corresponding magnetic structures are investigated using the Hubbard model within a mean field approximation.

\begin{figure*}[t!]
\centering
\includegraphics[width=13cm]{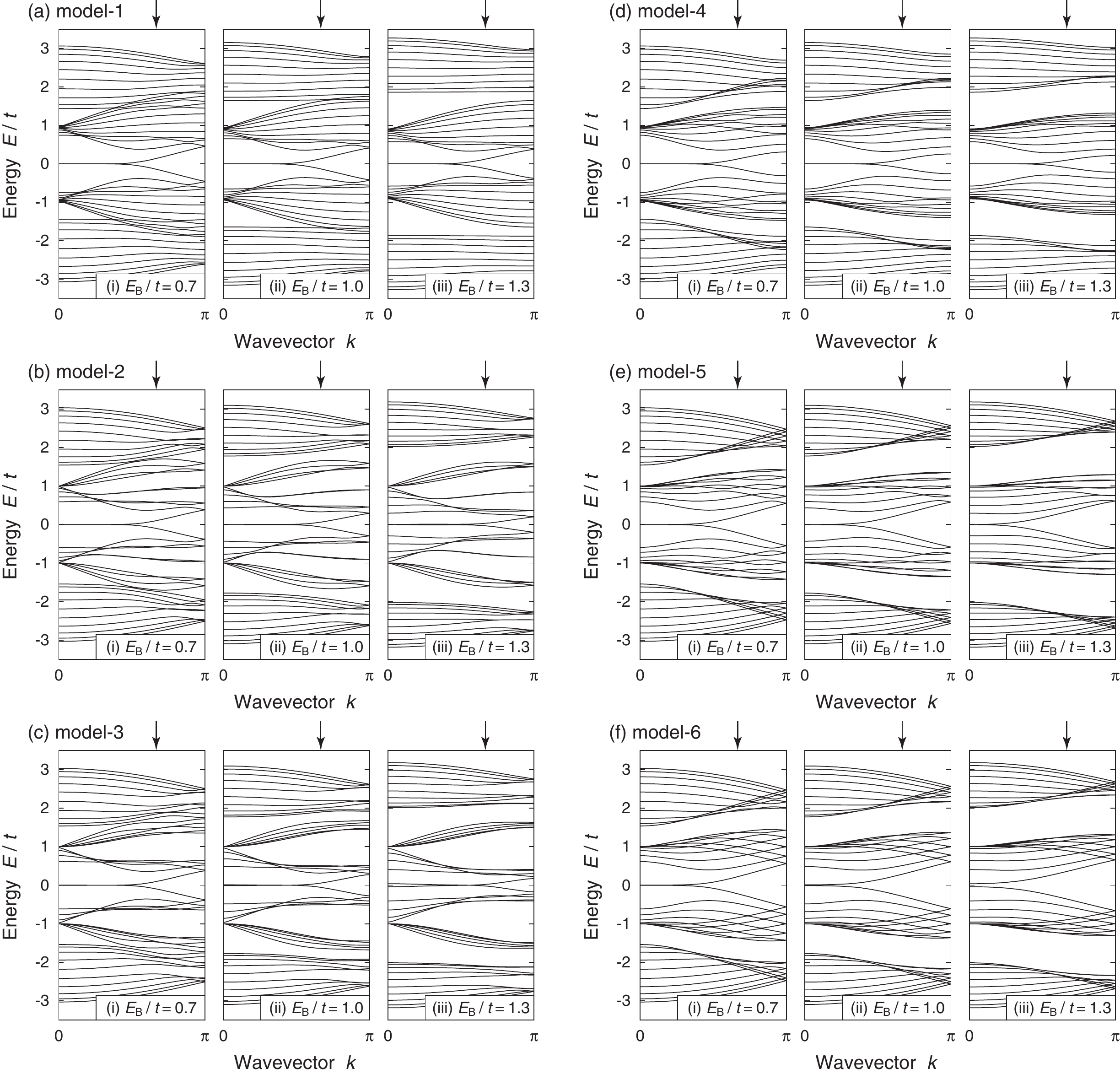}
\caption{
The band structures of BC$_2$N nanoribbons for $N=10$ and for the model-1 (a), model-2 (b), model-3 (c), model-4 (d), model-5 (e) and model-6 (f).
The results of $E_{\rm B}/t=0.7$ (i), 1.0 (ii) and 1.3 (iii) are presented in the left, middle and right panels, respectively.
The downward arrows indicate the position of $k=2\pi/3$.
}
\label{fg:N10-band}
\end{figure*}

\begin{figure*}[t!]
\centering
\includegraphics[width=13cm]{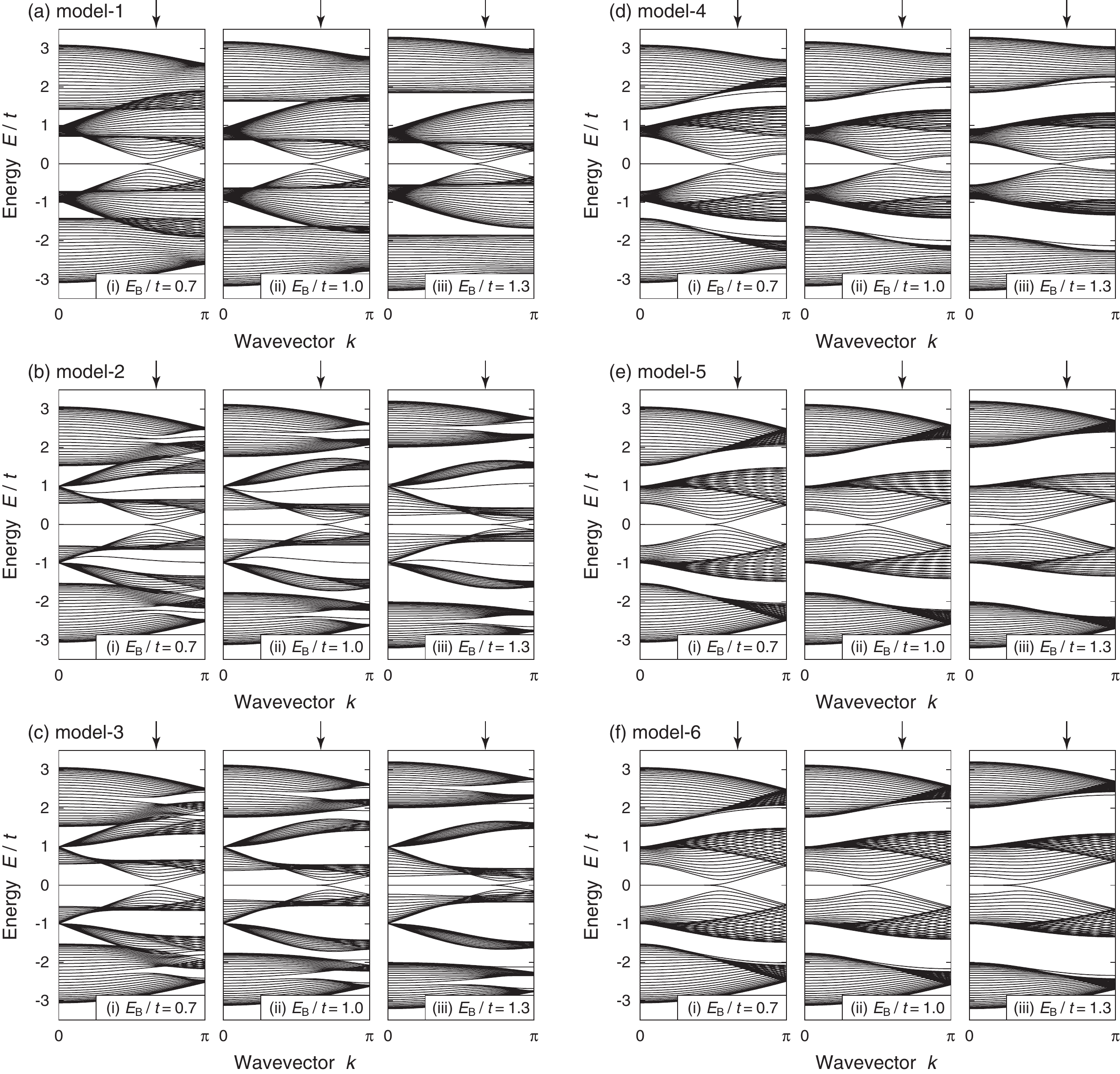}
\caption{
Same figure as Fig.\ 2 for $N=30$.
}
\label{fg:N30-band}
\end{figure*}

This paper is organized as follows:
In Sec.\ II, the models of BC$_2$N nanoribbons and the Hamiltonian of the system are presented.
The spin independent results within the tight binding model and the spin dependent results within the Hubbard models are  presented in Sec.\ III A and III B, respectively.
The discussion and short summary of the paper are given in Sec.\ IV  and V, respectively.
In Appendix A, the results with B-B and N-N bondings are summarized.

\section{Models and Methods}

We shall consider six different structures of BC$_2$N nanoribbons with zigzag edges as shown in Fig.\ \ref{fg:RibbonStructure}.
In this figure, B and N atoms, respectively, are indicated by the black and white circles, and C atoms are located the empty vertexes.
It should be noted that atoms are arranged as B-C-N-C along the zigzag line in these BC$_2$N nanoribbons. 
The dotted rectangles represent the unit cells and the shaded regions represent the unit cell of BC$_2$N nanoribbons and the shaded regions represent BC$_2$N sheets which consist of the nanoribbons. 
Let $N$ be a number of the zigzag lines. 
In this study, we shall restrict ourself to consider even $N$ cases for simplicity.
Because these bondings are not energetically favorable,  we do not consider the BC$_2$N nanoribbons with B-B or N-N bondings.
The results of such nanoribbons are summarized in Appendix A and the discussion about this issue is given in Sec.\ II C.

The Hamiltonian of the system within the tight binding model of $\pi$-electrons is given by
\begin{equation}
{\cal H}= \sum_{i} E_i c_{i}^\dagger c_{i} - \sum_{\langle i,j \rangle} t_{i,j} c_{i}^\dagger c_{j},
\label{eq:TB-Hamiltonian}
\end{equation}
where $E_i$ is an energy of $\pi$ electron at the site $i$, $c_{i}^\dag$ and $c_{i}$ are the creation and annihilation operators of electrons at the lattice site $i$, respectively, $\langle i,j\rangle$ stands for summation over the adjacent atoms and $t_{i,j}$ is the hopping integral of $\pi$ electrons from $j$th atom to $i$th atom.
$E_i$ are the site energies, $E_{\rm B}$, $E_{\rm C}$ and $E_{\rm N}$, at the B, C and N sites, respectively.
Following to the Yoshioka {\it et al}., we shall assume that the hopping integrals are constant regardless of the atoms, $t_{i,j}\equiv t$, $E_{\rm N}=-E_{\rm B}$ and $E_{\rm C}=0$. \cite{Yoshioka2003jpsj}
In Ref.\ [\onlinecite{Yoshioka2003jpsj}], $E_{\rm B}/t$ has been estimates as $0.8 \sim 0.9$.
However, the hopping integral, $t=3.0$ eV, used in Ref.\ [\onlinecite{Yoshioka2003jpsj}] seems to be overestimated compared with that determined by the first principles calculations within the local density approximation (LDA), $t=2.6\sim2.7$ eV.\cite{Yazyev2008prl}
On the other hand, the Shubunikov-de Haas measurements\cite{Zhang2005Nature} and the first-principles $GW$ calculations\cite{Trevisanutto2008prl,Yang2009prl} gave much larger values such as $t=3.4\sim3.5$ eV.
Furthermore, the site energy, $E_{\rm B}$, also have ambiguity.
In BN sheet, the energy gap is given by $2E_{\rm B}$.
The band gap was obtained 3.9 eV and 5.4 eV by the first-principles calculations within LDA and $GW$ approximation.\cite{Blase1995prb}
Ribeiro and Peres estimated as $t=2.33$ eV and $E_{\rm B}=1.96$ eV by fitting the  to the band structure of BN sheet within the tight binding model to that of the first-principles calculation within the generalized gradient approximation (GGA).\cite{Ribeiro2011prb}
Zheng {\it et al}.\ proposed that $t=2.621$ eV, $E_{\rm B}=1.894$ eV and the overlap integral $s=0.0154$ eV by fitting to the band structure of BN nanoribbon within the tight binding model to that of the first-principles calculation within LDA. \cite{Zheng2009jpsj}
But, $E_{\rm B}$ can vary about 2 eV depending on their environment.\cite{Kaneko2012arXiv-alternation}
To investigate the dependence of $E_{\rm B}$ on the electronic properties, therefore, we shall consider $E_{\rm B}/t=0.7$, $1.0$ and $1.3$.

In order to discuss the magnetism in BC$_2$N nanoribbons, we shall use the Hubbard model.
The Hamiltonian of the Hubbard model can be obtained by including the on-site Coulomb interaction to Eq.\ \eqref{eq:TB-Hamiltonian}:
\begin{equation}
{\cal H}= \sum_{i,\sigma} E_i c_{i,\sigma}^\dagger c_{i,\sigma}
- \sum_{\langle i,j \rangle} t_{i,j} c_{i,\sigma}^\dagger c_{j,\sigma}
+ \sum_{i} U_{i} n_{i,\uparrow}n_{i,\downarrow},
\label{eq:Hubbard-Hamiltonian}
\end{equation}
where $U_i$ is the on-site Coulomb interaction at $i$th site and $n_{i,\sigma}=c^\dag_{i,\sigma}c_{i,\sigma}$ ($\sigma=\uparrow$, $\downarrow$) is the number operator of electron at $i$t site with spin $\sigma$ .
In this study, we shall assume that $U_i$ is also constant, $U$.\cite{Kaneko2012arXiv-alternation}
We adopted the mean field approximation for this Hamiltonian in order to solve the problem.
The spin expectation values at $i$th sites are defined as $s_{i,z} =(\langle n_{i,\uparrow}\rangle - \langle n_{i,\downarrow}\rangle)/2$

The range of magnitude of $U$ is characterized by that of graphene.\cite{Yazyev2008prl,Yazyev2011prb}
In {\it trans}-polyacetylene which can be regarded as one-dimensional limit of graphene nanoribbons with zigzag edges, $U$ was estimated as 3.0 eV.\cite{Kuroda1984SolidStateCommun,Thomann1985prb}
According to the first-principles calculations within the local spin density approximation (LSDA) and GGA, $U/t$ was estimated as 0.9 and 1.3, respectively.\cite{Pisani2007prb,Gunlyche2007}
The magnetic structures in BCN nanoribbons within LSDA can be well reproduced in the Hubbard model. \cite{Kaneko2012arXiv-alternation}
In the previous study, $0\leq E_{\rm B}/t\leq 2$ and $0\leq U/t\leq 2$ were used and the results within LSDA can be well interpreted.
In this paper, we shall consider $0\leq U/t\leq 2$.

\section{Results and Discussion}

\subsection{Spin independent results}

In Fig.\ \ref{fg:N10-band}, the dependence of $E_{\rm B}$ on band structures of BC$_2$N nanoribbons for $N=10$ and for the model-1 (a), the model-2 (b), the model-3 (c), the model-4 (d), the model-5 (e) and the model-6 (f) are shown.
The results of $E_{\rm B}/t=0.7$, 1.0 and 1.3 are presented in the left (i), middle (ii) and right panels (iii), respectively.
The downward arrows indicate the position of $k=2\pi/3$, i.e., the projection of the K point of two dimensional graphene to the one dimensional system.
For $E_{\rm B}/t=0.7$ cases, we have found the flat bands at $E=0$ in the all BC$_2$N nanoribbons.  
The length of the flat bands in wavevector space seems depending on the atomic arrangements.
For $E_{\rm B}/t=1.0$ cases shown in the middle panels, we have observed the flat bands at $E=0$ but the length of the flat bands have been changed in some models, too.
With increasing  $E_{\rm B}$, the length of the flat bands increases for the model-2 and -3 nanoribbons, while the length decreases for the model-5 and -6 nanoribbons.  
On the other hand, for the model-1 and -4 nanoribbons, the length of the flat bands is not sensitive to the magnitude of $E_{\rm B}$. 
For $E_{\rm B}/t=1.3$ cases, these changes in band structures can be clearly seen.
For the model-2 and -3 nanoribbons, the length of the flat bands increases but the degeneracies of flat bands are lifted.
On the other hand, for the model-5 and -6 nanoribbons, the flat bands become much shorter. 
These results might indicate that there are threshold of $E_{\rm B}$ to have the flat bands.

In order to see the width dependence of the band structures, we shall consider the much wider BC$_2$N nanoribbons.
In Fig.\ \ref{fg:N30-band}, the similar figures as Fig.\ \ref{fg:N10-band} for $N=30$ are present. 
We have found that there are flat bands at $E=0$ for all nanoribbons independent of $E_{\rm B}$.
Furthermore, we can clearly see that the length of the flat bands increases for the model-2 and -3 nanoribbons but decreases for the model-5 and -6 nanoribbons with increasing  $E_{\rm B}$.
The length of the flat bands in the model-1 and -4 are independent of $E_{\rm B}$.
Therefore, the length of the flat bands depends on the the atomic arrangement and $E_{\rm B}$, and there is not a threshold value for the presence of the flat bands.  
It should be emphasized that we have confirmed that the flat bands are absent if atoms are not arranged as B-C-N-C along the zigzag lines.
Therefore, we can conclude that B-C-N-C arrangement along the zigzag line is necessary to obtain the flat bands.

\begin{figure}[t!]
\centering
\includegraphics[width=6.5cm]{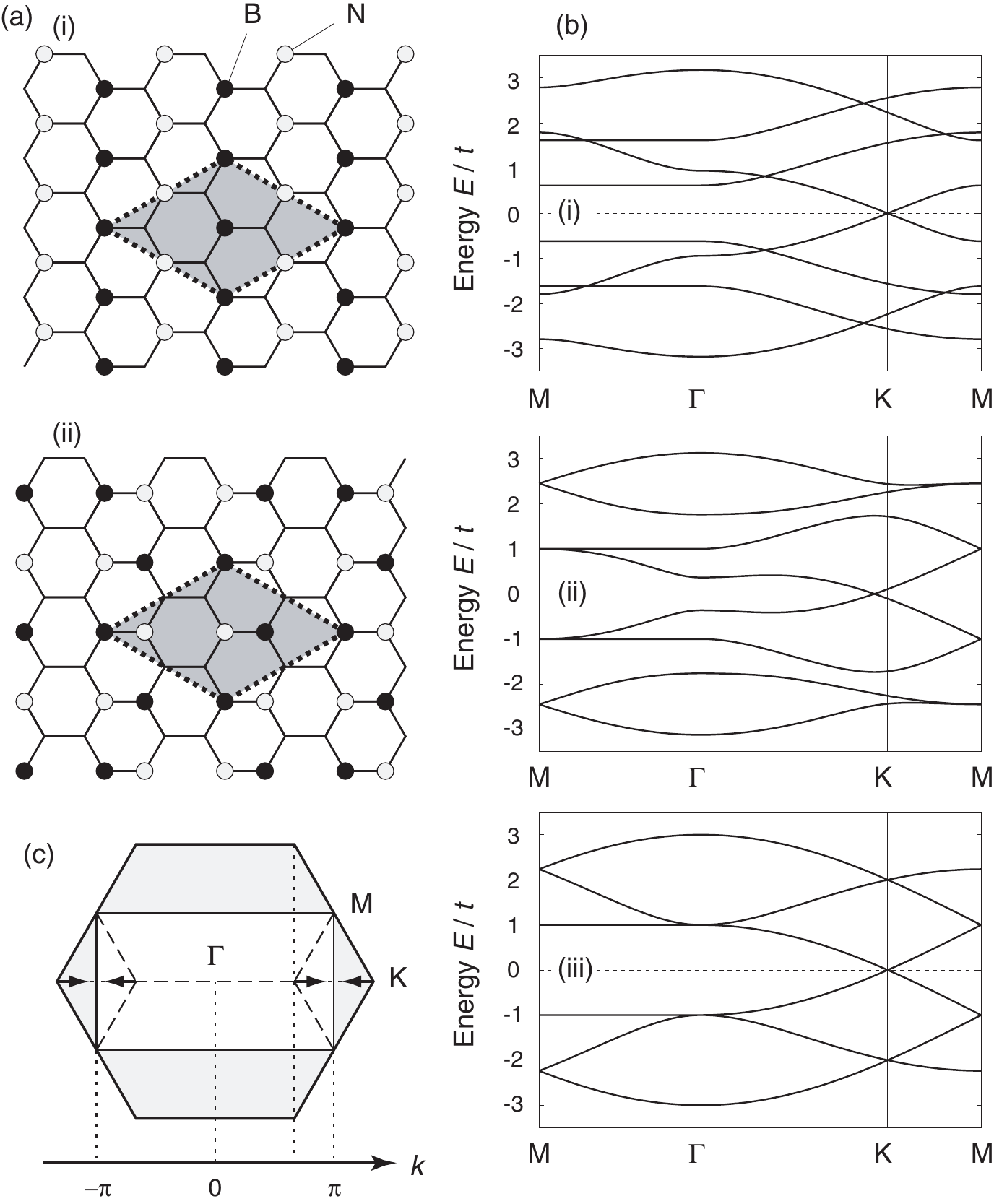}
\caption{
(a) Schematic illustration of BC$_2$N sheets presence of hexagonal symmetry.
BC$_2$N sheet shown in panel (i) consists the model-1 BC$_2$N nanoribbon and BC$_2$N sheet shown in panel (ii) consists the model-2 and 3 BC$_2$N nanoribbons.
(b) The band structures of BC$_2$N sheet for $E_{\rm B}/t=1.0$.
Panel (i)  and (ii) show the band structures of BC$_2$N sheets shown in panel (i) and (ii) of Fig.\ (a), respectively.
In panel (iii) corresponding band structure of graphene is shown.
(c) The Brillouin zone of the honeycomb lattice (the dashed lines) and its projection to the one dimensional Brillouin zone of zigzag nanoribbons.
}
\label{fg:Hex2D-band}
\end{figure}

\begin{figure}[t!]
\centering
\includegraphics[width=6.5cm]{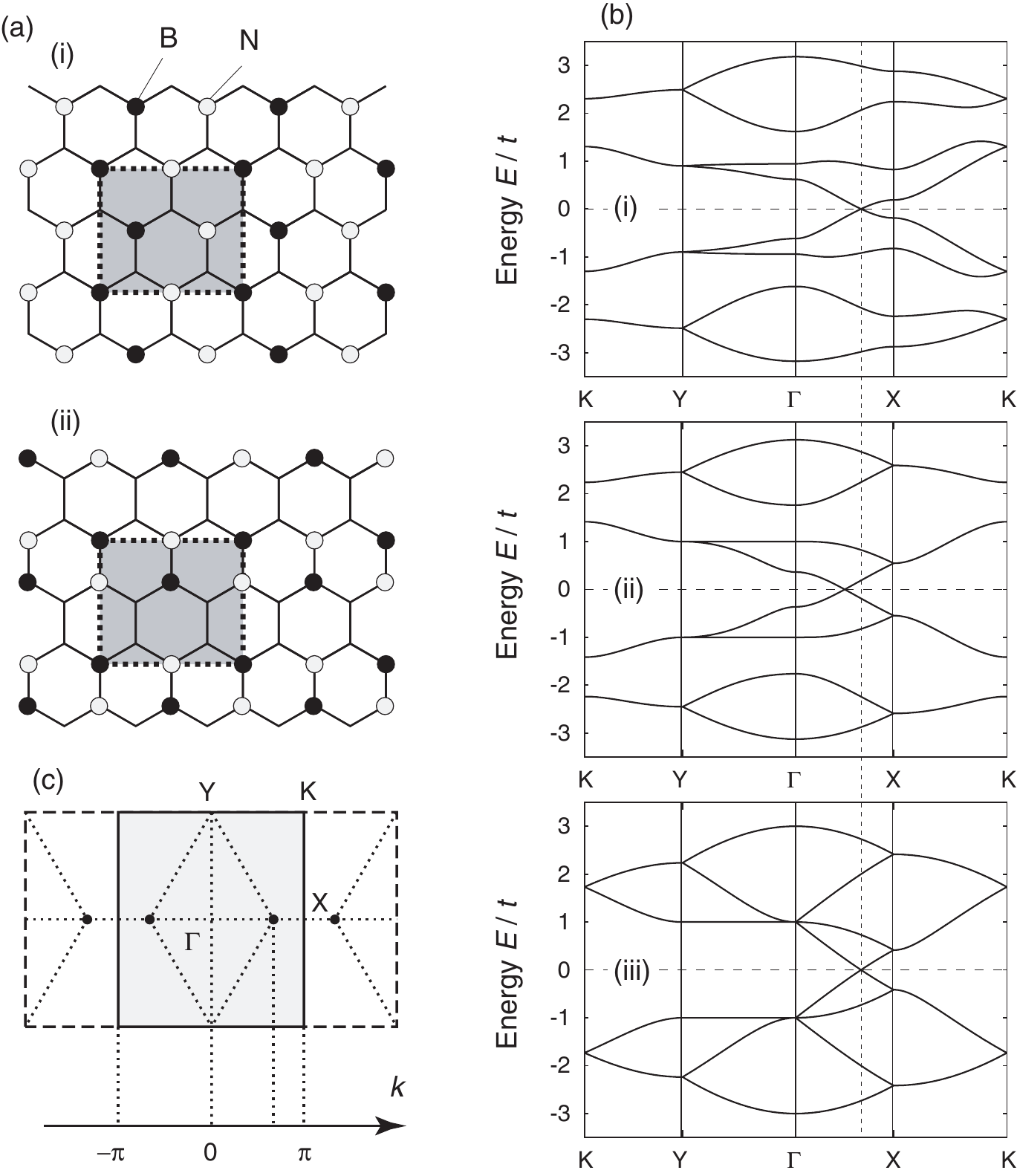}
\caption{
(a) Schematic illustration of BC$_2$N sheets absence of hexagonal symmetry.
BC$_2$N sheet shown in panel (i) consists the model-4 BC$_2$N nanoribbon and BC$_2$N sheet shown in panel (ii) consists the model-5 and 6 BC$_2$N nanoribbons.
(b) The band structures of BC$_2$N sheet for $E_{\rm B}/t=1.0$.
Panel (i)  and (ii) show the band structures of BC$_2$N sheets shown in panel (i) and (ii) of Fig.\ (a), respectively.
In panel (iii) corresponding band structure of graphene is shown.
(c) The Brillouin zone of the rectangular lattice.
}
\label{fg:Rec2D-band}
\end{figure}

\begin{figure*}[t!]
\centering
\includegraphics[width=14cm]{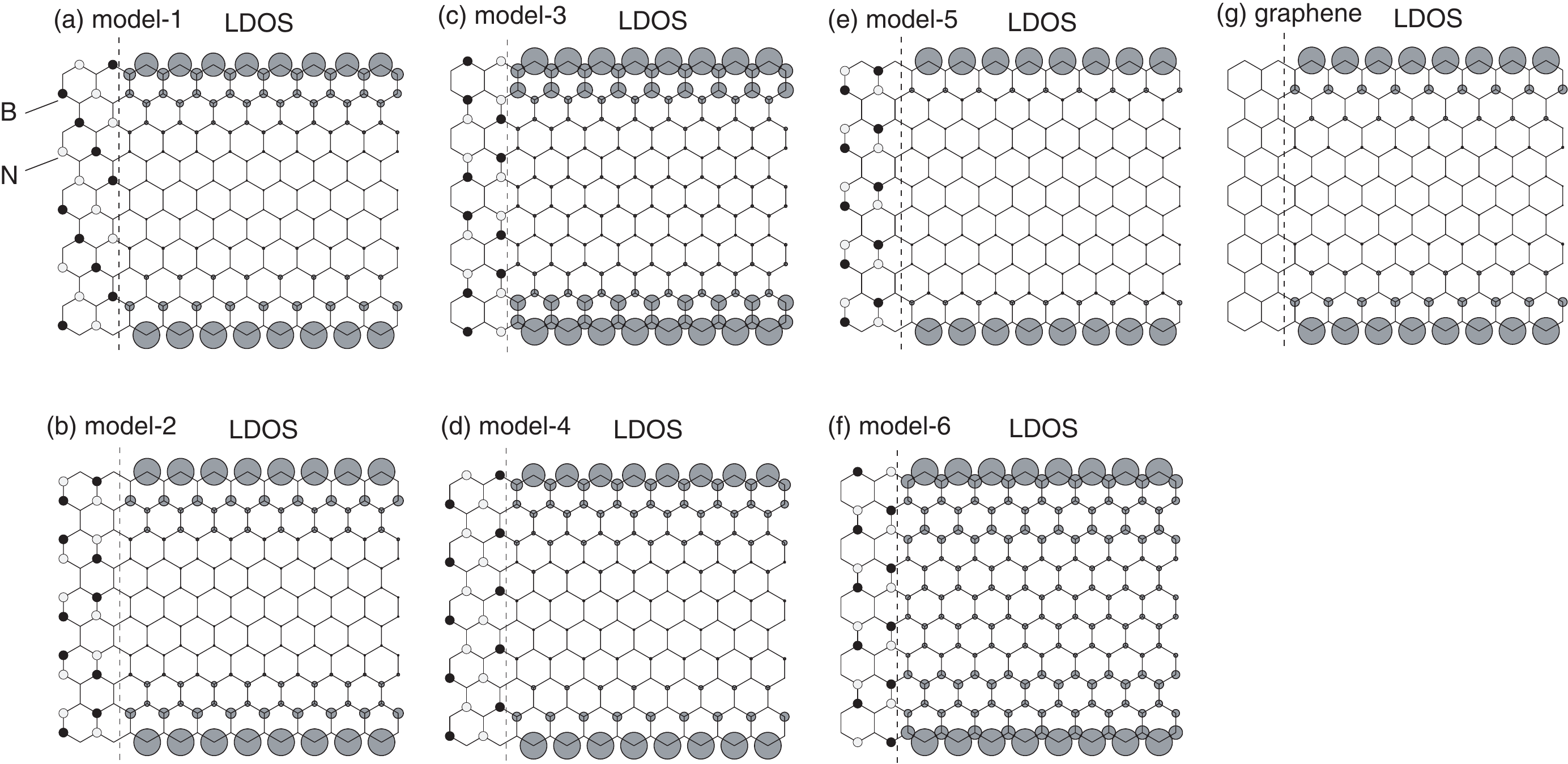}
\caption{
Calculated local density of states (LDOS) for $E_{\rm B}/t=1.0$ and $N=10$ with schematic illustrations for the model-1 (a), model-2 (b), model-3 (c), model-4 (d), model-5 (e) and model-6 (f).
As the reference, LDOS of graphene nanoribbons in doubled unit cell is presented in (g).
In the left side of each panel, the corresponding schematic illustration is presented.
The magnitude of LDOS is indicated by the area of the circles.
}
\label{fg:LDOS}
\end{figure*}

In order to understand the difference in the length of the flat bands in the wavevector space, we shall consider the band structure of BC$_2$N sheet.
In Figs.\ \ref{fg:Hex2D-band} (a), schematic illustrations of BC$_2$N sheets are present.
The shaded region indicates the unit cell for the band calculations.  
It is noted that the model-1 nanoribbon consists of BC$_2$N sheet shown in Fig.\ \ref{fg:Hex2D-band} (a)-(i) while the model-2 and -3 nanoribbons consist of BC$_2$N sheet shown in Fig.\ \ref{fg:Hex2D-band} (a)-(ii). 
Figure \ref{fg:Hex2D-band} (b)-(i) and (ii) show the band structures of BC$_2$N sheet for $E_{\rm B}/t=1.0$ shown in Fig.\ \ref{fg:Hex2D-band} (a)-(i) and (ii), respectively. 
In Fig.\ \ref{fg:Hex2D-band} (b)-(iii), the band structure of graphene in $2\times2$ supercell is presented as a reference.
For BC$_2$N sheet shown in Fig.\ \ref{fg:Hex2D-band} (a)-(i), the linear dispersion is realized at the K point.
This feature is quite similar to that of graphene but the group velocity around the Fermi level is slightly smaller than that of graphene.
On the other hand, for BC$_2$N sheet of Fig.\ \ref{fg:Hex2D-band} (a)-(ii), the position of linear dispersion deviates from the K point toward the $\Gamma$ point. 
The deviation of the linear dispersion increases with increasing  $E_{\rm B}$.  
This result is quite similar to the result calculated by Liu {\it et al}.\ using the first principles calculations.\cite{Liu1989prb}
As discussed below, the position of the Dirac points plays decisive role to determine the length of flat bands.

Figure \ref{fg:Hex2D-band} (c) shows the Brillouin zone of the honeycomb lattice and its projection to the one dimensional Brillouin zone of zigzag nanoribbons. 
The projection of the K point of the honeycomb lattice, i.e., the Dirac point of graphene and BC$_2$N sheet shown in Fig.\ \ref{fg:Hex2D-band} (a)-(i)  is located at $k=2\pi/3$. 
Then, the model-1 nanoribbons have the flat bands in $0\leq k\leq 2\pi/3$ which is same region as graphene nanoribbons.
When the Dirac points deviate from the K point toward the $\Gamma$ point as indicated by the arrows, the projection of the Dirac points shift from $k=2\pi/3$ toward $k=\pi$ as shown in the bottom of Fig.\ \ref{fg:Hex2D-band} (c).
So, the length of the flat bands increases.
As the result, the difference in the band structures of BC$_2$N sheets might lead to the difference in the length of the flat bands in wavevector space.

Next, we shall consider the BC$_2$N sheet shown in Fig.\ \ref{fg:Rec2D-band}.
The model-4 nanoribbon consists of BC$_2$N sheet are presented in Fig.\ \ref{fg:Rec2D-band} (a)-(i) while the model-5 and -6 nanoribbons consist of BC$_2$N sheet are shown in Fig.\ \ref{fg:Rec2D-band} (a)-(ii).
In Fig.\ \ref{fg:Rec2D-band} (a), the band structures of BC$_2$N sheets for $E_{\rm B}/t=1.0$ are shown in Fig.\ \ref{fg:N10-band} (b).
The band structure of graphene in the same unit cell is shown in Fig.\ \ref{fg:Rec2D-band} (b)-(iii). 
The corresponding Brillouin zone is indicated in Fig.\ \ref{fg:Rec2D-band} (c). 
The vertical dotted line in Fig.\ \ref{fg:N10-band} (b) indicates the position of the K point of honeycomb lattice.
For BC$_2$N sheet shown in Fig.\ \ref{fg:Rec2D-band} (a)-(i), the linear dispersion is located at the same point of the graphene.
As displayed in the lower part of Fig.\ \ref{fg:Rec2D-band} (c), the projection of the K point of honeycomb lattice is located at $k=2\pi/3$ in the one dimensional Brillouin zone of nanoribbons.
Therefore, the length of the flat bands in the wavevector space is same as those in the graphene nanoribbons.  
On the other hand, the point of linear dispersion shifts toward the $\Gamma$ point in BC$_2$N sheet shown in Fig.\ \ref{fg:Rec2D-band} (a)-(ii).
In this case, the position of the linear dispersion in the one dimensional Brillouin zone also shift toward $k=0$.
Then, the length of the flat bands of the model-5 and -6 BC$_2$N nanoribbons is shorter than those of graphene.
Furthermore, the length of flat bands of other BC$_2$N nanoribbons also vary in a similar manner as discussed Appendix A.
To conclude, we have found that the length of the flat bands in BC$_2$N nanoribbons is determined by the position of the Dirac points in the Brillouin zone.

Next, we move to the charge distribution in zigzag BC$_2$N nanoribbons.
In Fig.\ \ref{fg:LDOS}, the local density of states (LDOS) at $E=0$ for several structures with $E_{\rm B}/t=1.0$ are shown by the circles.
In this figure, the area of the circles is proportional to the magnitude of the LDOS at each site. 
In the left side of each panel, the corresponding schematic illustration is presented.  
As the reference, the LDOS at $E=0$ graphene nanoribbon is shown in Fig.\ \ref{fg:LDOS} (g). 
In the model-1 and -4 nanoribbons shown in Figs.\ \ref{fg:LDOS} (a) and (d), the upper and lower edge are different each other.
In the following, we shall call the upper (lower) edge as B, N side (C side) edge.
The electronic charge is localized at the BC$_2$N nanoribbons edges, showing the presence of the edge states. 
As discussed below, the edge states in BC$_2$N nanoribbons are different from those in conventional graphene nanoribbons.

\begin{figure}[b!]
\centering
\includegraphics[width=6.5cm]{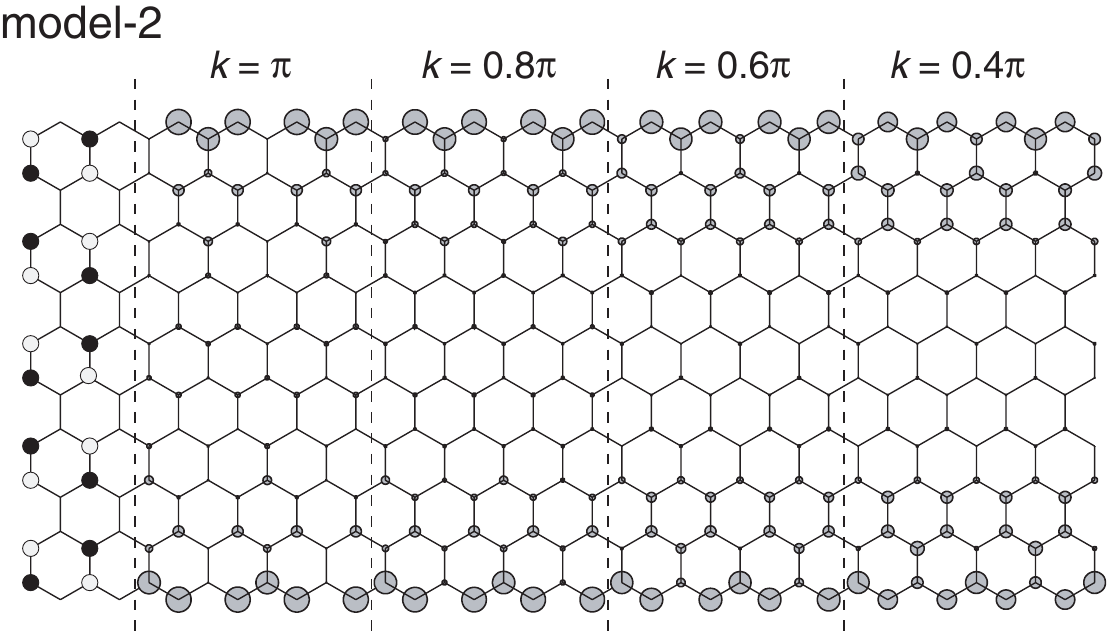}
\caption{
The charge distribution of the edge states with $E>0$ in the model-2 BC$_2$N nanoribbons for $N=10$ and $E_{\rm B}/t=1.0$.
The areas of the circles indicate the magnitude of the charge densities.
}
\label{fg:chg-model2}
\end{figure}

In the model-1, the charge distributions along the B, N side and C side edges are different each other. 
The charge distribution at the C side edge is similar to that at the conventional zigzag edge of graphene nanoribbons, while the charge of the edge states at the B, N side edge distributes the both sublattice sites.
Recently, Kaneko {\it et al}.\ have shown that the edge states in zigzag graphene nanoribbons are robust on the substitution of outermost C atoms with B and N atoms alternately.\cite{Kaneko2012arXiv-alternation}
However, such substitution causes change in charge distribution, i.e., the sublattice structure is broken.\cite{Kaneko2012arXiv-alternation}
The edge states at the B, N side edge is similar to those discovered by Kaneko {\it et al}.\cite{Kaneko2012arXiv-alternation}
In the model-2 nanoribbon, the charge distribution of the edge states is similar to that of graphene nanoribbons but the sublattice structure is broken inside the nanoribbons.  In the model-3 nanoribbon, the charge distributes over both sublattice sites, showing the similarity of those discovered by 
Kaneko {\it et al}.\cite{Kaneko2012arXiv-alternation}
The charge distribution in the model-4 BC$_2$N nanoribbon is quite similar to that in the model-1 BC$_2$nanoribbon.
The charge distributions in the model-5 and -6 BC$_2$N nanoribbons are similar to those in the model-2 and -3 BC$_2$N nanoribbons, respectively.
However, the LDOS inside of nanoribbons in the model-5 and -6 are smaller than those in the model-2 and -3 nanoribbons.
This difference is caused by the difference in the length of the flat bands in the wavevector space.

\begin{figure*}[t!]
\centering
\includegraphics[width=14cm]{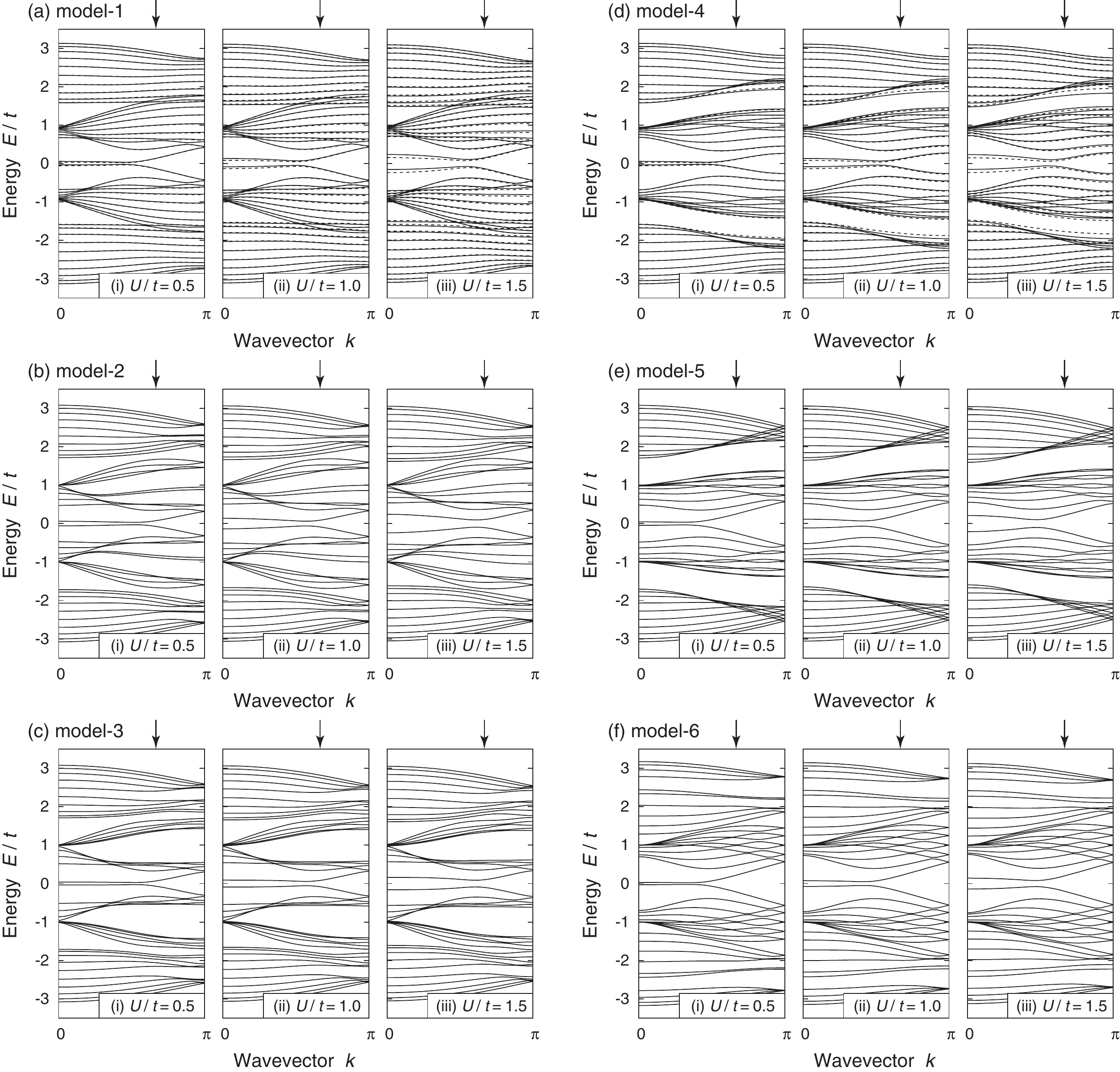}
\caption{
The spin dependent band structures of BC$_2$N nanoribbons for $N=10$ and $E_{\rm B}/t=1.0$, and for the model-1 (a), model-2 (b), model-3 (c), model-4 (d), model-5 (e), model-6 (f).
The results of $U/t=0.5$ (i), 1.0 (ii) and 1.5 (iii) are presented in the left, middle and right panels, respectively.
}
\label{fg:spin-N10-band}
\end{figure*}

In the model-2 nanoribbons, there are other edge states even $E\neq 0$.
As shown in Fig.\ \ref{fg:N10-band} (b) and \ref{fg:N30-band} (b), we have found width independent energy bands around $E/t\sim\pm1$ in $k/\pi\sim 0.2-1$. 
The wavevector dependence of the charge distribution for $N=10$ $E_{\rm B}/t=1.0$ of the edge states with $E>0$ is shown in Fig.\ \ref{fg:chg-model2}. 
In this figure, the areas of the circles indicate the magnitude of the charge densities.
We can clearly see the formation of the edge states even $E\neq0$.
At $k=\pi$, the electrons are mainly accumulated in the outermost C atoms and the next outermost B atoms, while the electronic charge is absent at the next outermost N atoms. 
With decreasing of $k$, the electronic charge at the next outermost N atoms increases, resulting in the decrease in the energy.
For $E<0$ states, on the other hand, the electrons  are mainly accumulated in the outermost C atoms and the next outermost N atoms.

\begin{figure*}[t!]
\centering
\includegraphics[width=14cm]{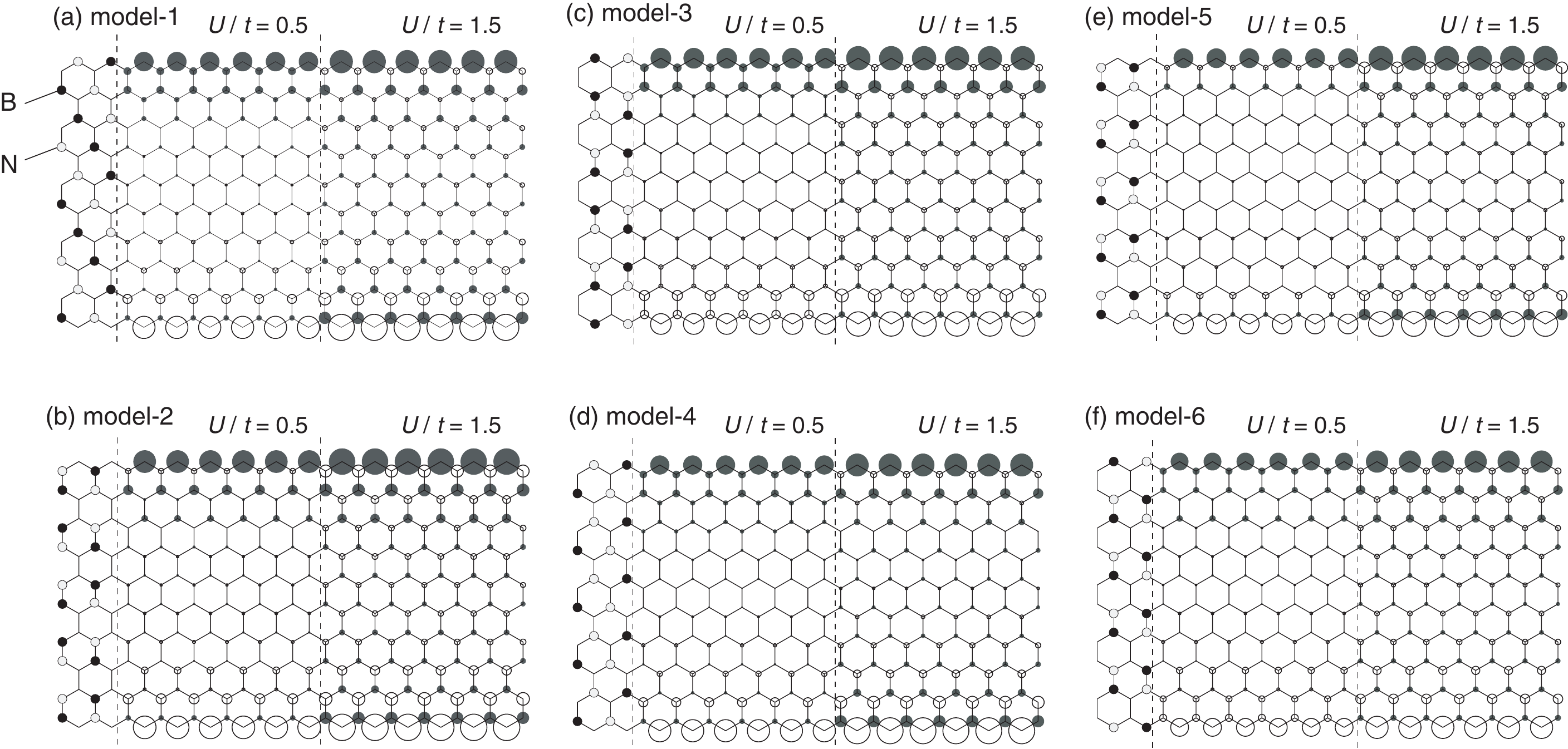}
\caption{
Calculated spin densities for $E_{\rm B}/t=1.0$ and $N=10$ with schematic illustrations (left side), and for the model-1 (a), model-2 (b), model-3 (c), model-4 (d), model-5 (e) and model-6 (f).
The results for $U/t=0.5$ and 1.5 are presented in the middle and right side in each panel.
The up (down) spin densities are denoted by the solid (open) circles whole areas are proportional to the magnitudes of the spin densities.
}
\label{fg:SPIN}
\end{figure*}

\subsection{Spin dependent results}

Next, we shall consider the magnetism in BC$_2$N nanoribbons.
Figure \ref{fg:spin-N10-band} shows calculated band structures of BC$_2$N nanoribbons for $E_{\rm B}/t =1.0$ and $N=10$.
The results of $U/t=0.5$, 1.0 and 1.5 are presented in the left, middle and right panels, respectively.
In Figs. \ref{fg:spin-N10-band} (a) and (d), the solid and dashed lines represent the energy bands of up and down spins, respectively.
It should be noted that the band structures of up and down spins of the model-2, -3, -5 and -6 nanoribbons shown in Fig.\ \ref{fg:spin-N10-band} (b), (c), (d) and (c) are double degenerate.

We found that the degeneracy of flat bands are lifted due to the presence of the Coulomb interaction independent of the atomic arrangement.
In the model-2 nanoribbons, the degenerate flat bands of $E\neq 0$ are also lifted.
These BC$_2$N nanoribbons are semiconductor with direct gap.
The band gap increases with increasing  $U/t$.
The position of valence tops and conduction bottoms are located around $k=0.4\pi \sim 0.7\pi$ depending on the atomic arrangement, showing good agreement with the position of projection of the Dirac point.

Corresponding spin density distributions for $U/t=0.5$ (middle) and $1.5$ (right side) are present in Fig.\  \ref{fg:SPIN} with their schematic illustrations (left side).
In this figure, the up (down) spin densities are denoted by the solid (open) circles whose areas are proportional to the magnitudes of the spin densities.
In the model-1 nanoribbon, the ferrimagnetic order is observed at the C terminated edge while the ferromagnetic order is observed in B and N terminated edge for $U/t=0.5$.
For $U/t=1.5$, on the other hand, the ferromagnetic orders are observed at the both edges.
In the model-2 nanoribbon, the ferrimagnetic orders are observed at the both edges independent of the magnitude of $U$.
However, the magnetic structures in the model-3 depend on the magnitude of $U$ while the model-2 and -3 nanoribbons are consisted with same BC$_2$N sheet.
We observed ferromagnetic order at the both edges for $U/t=0.5$ but it turns into ferrimagnetic with increasing in $U$. 
Therefore, we found that the magnetic structures in BC$_2$N nanoribbons depend on both the atomic arrangements and the magnitude of $U$.

It should be noted that the ferromagnetic orders were observed when the outermost site are occupied by B and N atoms alternately and $U/t=0.5$.
At these edges, LDOS shown in Fig.\ \ref{fg:LDOS} distributed over both sublattice sites.
For the ferrimagnetic orders even for $U/t=0.5$, on the other hand, LDOS at the next outermost sites are absent.
Therefore, the charge distributions seem to determine the magnetic structure in BC$_2$N nanoribbons.
These behaviors are quite resemble to the previously reported magnetic orders in BCN nanoribbons 
where the outermost C atoms of graphene nanoribbons are replaced by B and N atoms alternately.

To confirm this conjecture, we shall see the spin distributions in the model-4, 5 and 6 nanoribbons.
The magnetic structure of the model-4 nanoribbons for $U/t=0.5$ is the ferrimagnetic at C side edge but ferromagnetic at B,N side edge.
For $U/t=1.5$, the magnetic structures at both edges are the ferrimagnetic.
The magnetic structures of the model-4 nanoribbon shown in Fig.\ \ref{fg:SPIN} (d) are quite resemble to those of the model-1, showing good agreement of the conjecture.
In the model-5 nanoribbons, we observed the ferrimagnetic structures independent of the magnitude of $U$.
In this nanoribbons, the outermost sites are occupied by C atoms and LDOS at the next outermost sites are absent. 
In the  model-6 nanoribbons, on the other hand, the ferromagnetic orders are observed for $U/t=0.5$ but they turn into ferrimagnetic with increasing  $U$.
The outermost sites of the model-6 nanoribbons are occupied with B and N atoms and the electronic charges distribute over both sublattice sites.
We found that the magnetic structures in the model-5 and -6 nanoribbons satisfy the conjecture.

\begin{figure*}[t!]
\centering
\includegraphics[width=14cm]{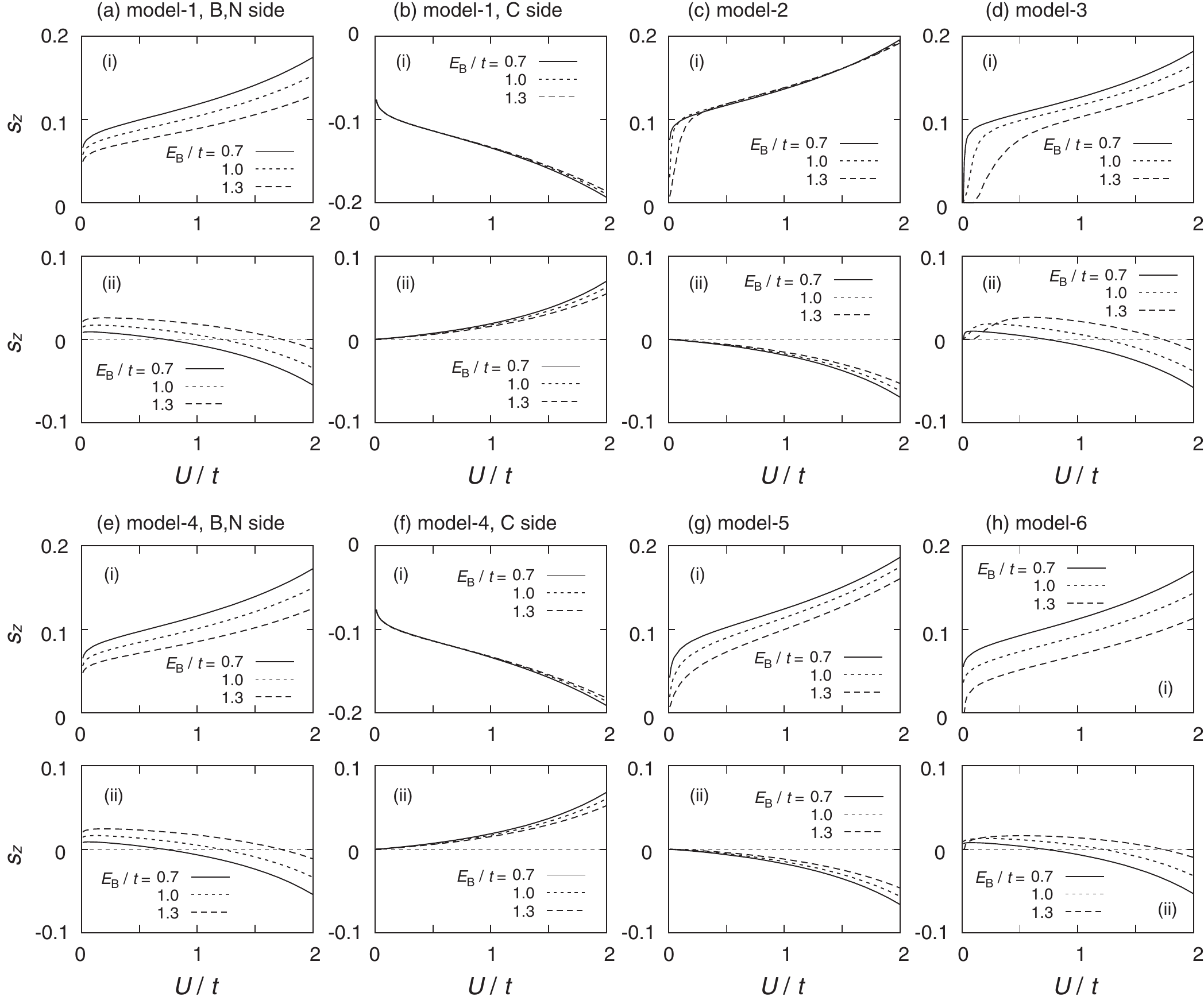}
\caption{
$U$ dependence of spin expectation values at the outermost sites (i) and the next outermost sites (ii) at B,N side in the model-1 (a),  C side in the model-1 (b),  model-2 (c), model-3 (d), B,N side in the model-4 (e),  C side in the model-4 (f),  model-5 (g) and model-6 (h) BC$_2$N nanoribbons.
The solid, dotted and dashed lines represent the results for $E_{\rm B}/t=0.7$, 1.0 and 1.3, respectively.
}
\label{fg:SPIN-1-2}
\end{figure*}

We shall consider $E_B$ and $U$ dependence of magnetic structures in order to to clarify when change in magnetic structure takes place.
The dependence of $U$ of the spin densities are summarized in Fig.\ \ref{fg:SPIN-1-2}.
In this figure, spin expectation values at the outermost sites and the next outermost sites are presented in the top and bottom panels, respectively.
The results of $E_{\rm B}/t=0.7$, 1.0 and 1.3 are represented by the solid, dotted and dashed lines, respectively.
It should be emphasized that obtained spin expectation values in B and N atom are same as we have seen in Fig.\ \ref{fg:SPIN}.
Therefore, we plotted the spin expectation values in C atoms and B atoms.

At the B, N side of model-1 nanoribbons, we obtained positive $s_z$ independent of $E_{\rm B}$ at the outermost and next outermost sites for sufficiently small $U$.
With increasing  $U$, $s_z$ at the outermost site monotonically increases while that at the next outermost site monotonically decrease, showing that the magnetic structures change from the ferromagnetic into ferrimagnetic order depending on $E_{\rm B}$.
At the C-side of model-1, on the other hand, $s_z$ of the outermost and next outermost sites are negative and zero, respectively, for sufficiently small $U$. 
With increasing in $U$, $s_z$ of the outermost monotonically decreases but that of next outermost sites increase, i.e., the ferrimagnetic order is observed.
Since B and N atoms are arranged in the sublattice sites which do not belong to the outermost sites, the $E_{\rm B}$ dependence of $s_z$ is small at the C-side compared with those at the B, N-side.
In the model-4 nanoribbons, calculated $s_z$ at both edges show quite similar dependence on $U$ to those in the model-1 nanoribbons since the electronic properties of these nanoribbons around $E=0$ resemble between them.

In the model-2 nanoribbons, $s_z$ are almost independent of $E_{\rm B}$ except $U/t<0.2$ regime.
As we discussed before, the degeneracy of flat bands are lifted with inceasing of $E_{\rm B}$, resulting in decreasing of strength of the electron correlation effect.
Therefore, $s_z$ becomes smaller with increasing  $E_{\rm B}$.
Similar effect can be found in the model-3 nanoribbons.
Especially, $s_z$ vanishes for $U/t<0.1$ in the model-3 nanoribbons with $E_{\rm B}/t=1.3$.
With increasing  $U$, $s_z$ at the next outermost site once increases but turns to decrease to negative value at certain value of $U$.
Then, the ferrimagnetic order is constructed.
In the model-5 and -6 nanoribbons, obtained $s_z$ shows similar dependence on $U$ to those in the model-2 and -3 nanoribbons, respectively, but the magnitudes of $s_z$ becomes smaller since the length of flat bands are shorter than those in the model-5 and -6 nanoribbons.
As we mentioned before, the degeneracies are recovered with increase of the ribbon width, suggesting that the strength of electron correlation effect should increases.
It should be emphasized that we confirmed that such strong $E_{\rm B}$ dependence for small $U$ will decrease with increasing  the ribbon width.

\begin{figure*}[t!]
\centering
\begin{tabular}{lr}
\begin{minipage}{0.68\hsize}
\includegraphics[width=10cm]{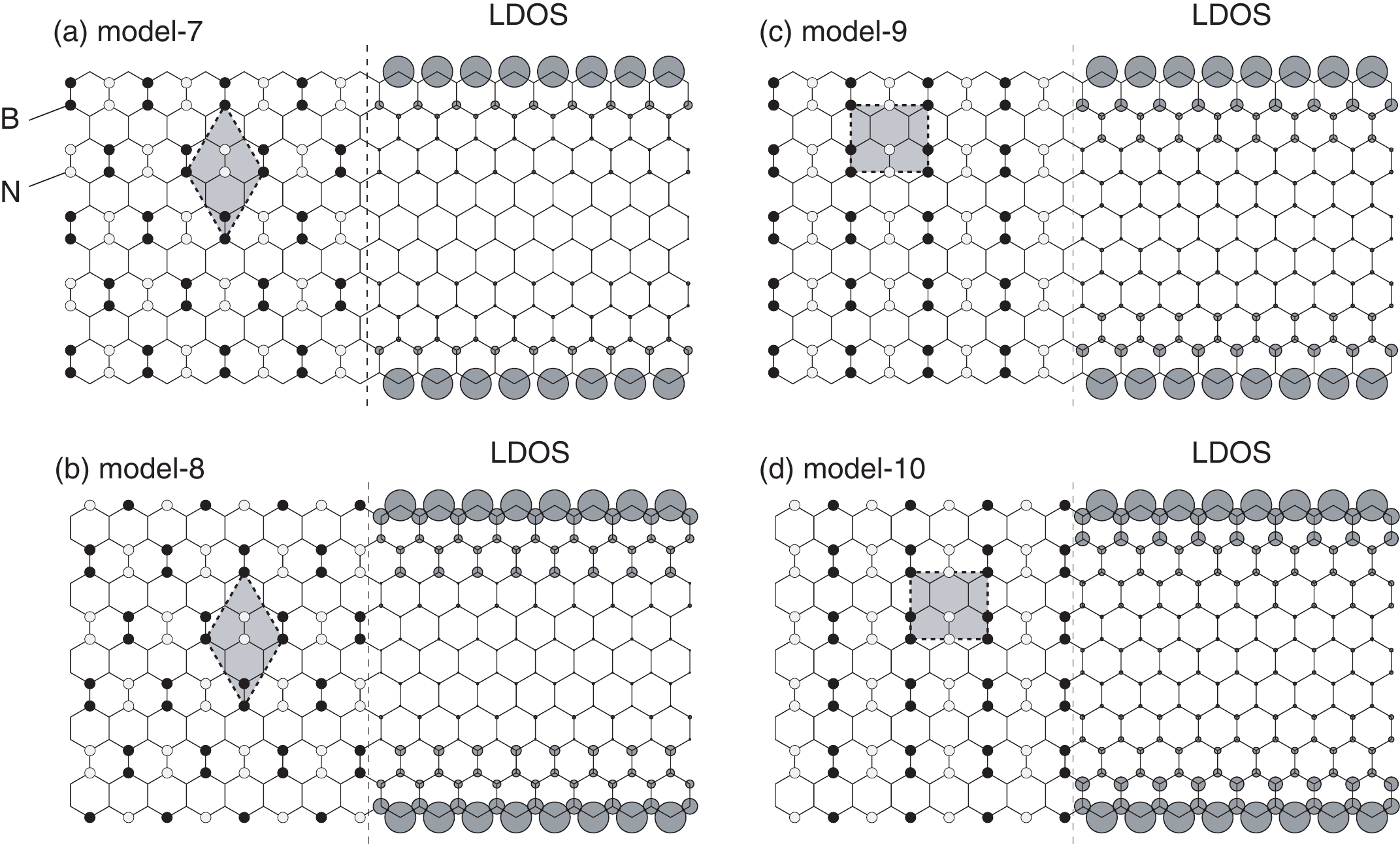}
\end{minipage}
&
\begin{minipage}{0.3\hsize}
\caption{
The models of BC$_2$N nanoribbons handled in the appendix (left sides) and calculated LDOS at $E=0$ for $E_{\rm B}/t=1.0$ (right sides).
The dotted rectangles represent the unit cells and the shaded regions represent the unit cell of BC$_2$N sheets.
}
\label{fg:add-Structure-LDOS}
\end{minipage}
\end{tabular}
\end{figure*}

\section{Discussion}

We did not consider BC$_2$N nanoribbons in which atoms are arranged as B-C-N-C along zigzag lines with B-B and N-N bonds.
The results of BC$_2$N nanoribbons with B-B and N-N bonds are summarized in Appendix A.
We confirmed that such BC$_2$N nanoribbons also have flat bands at the Fermi level but the length of the flat bands of such BC$_2$N nanoribbons also different from that in conventional zigzag graphene nanoribbons.
It should be noted that the change is the length of the flat bands can be understood as the change in the Dirac point.
The magnetic structures in these BC$_2$N nanoribbons were calculated using the Hubbard model within the mean field approximation.
For the magnetic structures in these BC$_2$N nanoribbons, the atomic arrangement of the outermost sites and the ratio of the site every to the on-site Coulomb interaction also play decisive role as discussed above.

\begin{figure*}[t!]
\centering
\includegraphics[width=14cm]{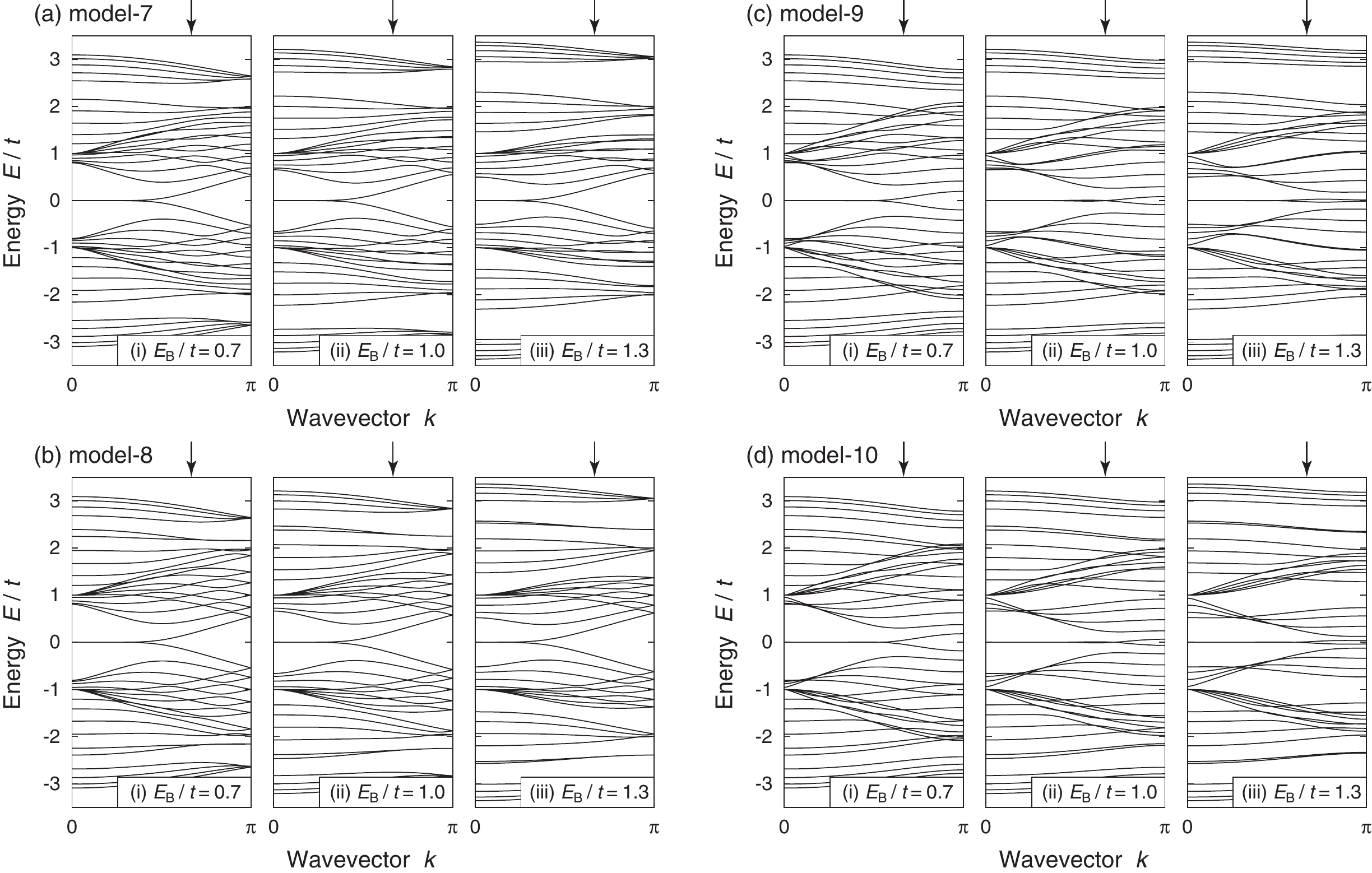}
\caption{
The band structures of BC$_2$N nanoribbons for $N=10$ and for the model-7 (a), model-8 (b), model-9 (c) and model-10 (d).
The results of $E_{\rm B}/t=0.7$ (i), 1.0 (ii) and 1.3 (iii) are presented in the left, middle and right panels, respectively.
The downward arrows indicate the position of $k=2\pi/3$.
}
\label{fg:add-N10-band}
\end{figure*}

\begin{figure*}[t!]
\centering
\includegraphics[width=14cm]{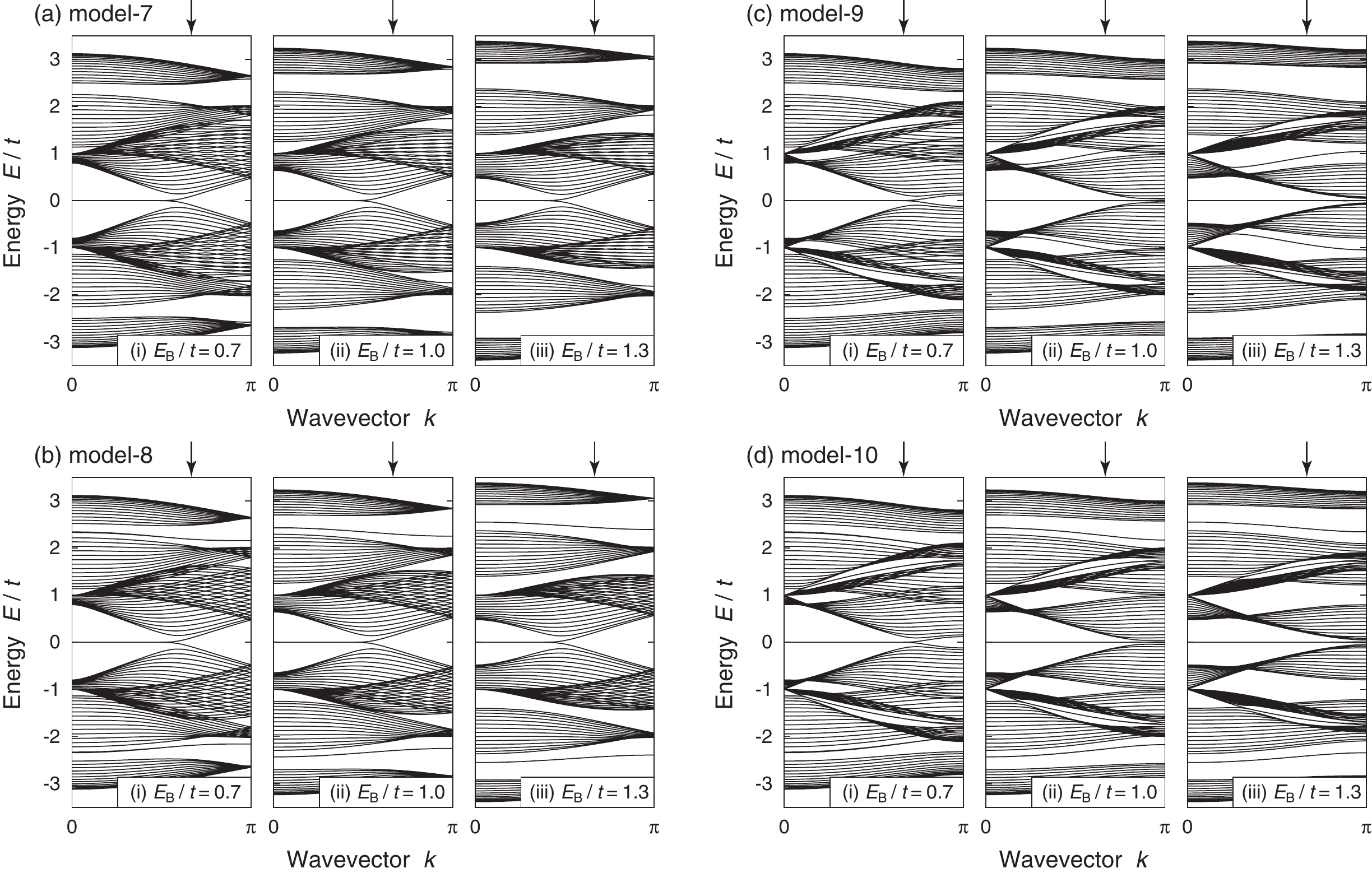}
\caption{
Same figure as Fig.\ \ref{fg:add-N10-band} for $N=30$.
}
\label{fg:add-N30-band}
\end{figure*}

Dependence of atomic arrangement on the energetics of BC$_2$N sheet has been investigated by Liu {\it et al}.\ and Azevedo using the first-principles calculations.\cite{Liu1989prb,Azevedo2005EurPhysJB}
According to their study, BC$_2$N sheets stabilize with increasing  B-N and C-C bondings, suggesting that BN  islands and graphene islands hybridized structure is most stable. 
Such tendency was confirmed in the arbitrary stoiciometric BCN, i.e., B$_x$C$_y$N$_z$ sheets, by the several authors.\cite{Mazzoni2006prb,Azevedo2006epl,Manna2011jpcc}
Therefore, the BC$_2$N nanoribbons treated in this paper are not stable.

While the atomic structures of synthesized BC$_2$N was not identified, Watanabe {\it et al}.\ reported the bonding characteristic of BC$_2$N tho films by means of x-ray photoelectron spectroscopy measurement of chemical shift of $1s$ electrons.\cite{Watanabe1996apl}
According to their study, BC$_2$N thin films have significant B-C and C-N bondings, i.e.,  all B, C and N atoms bond with one another and mixed atomically.\cite{Watanabe1996apl}
Therefore, the synthesized BC$_2$N might have the atomic arrangement considered in this paper.
The synthesized graphitic BC$_2$N showed p-doped semiconducting character with band gap about 2 eV by STM and photoluminescence measurement.\cite{Watanabe1996prl}
However, BC$_2$N sheets considered in this paper are metallic as shown in Figs.\ \ref{fg:Hex2D-band} (b) and \ref{fg:Rec2D-band} (b).
The tight binding model cannot describe the charge transfer as discussed in Ref.\ [\onlinecite{Yoshioka2003jpsj}] due to its simplicity.
The effect of charge transfer might become important in BCN nanoribbons.\cite{Kaneko2012arXiv-GSE} 
Therefore, the discussion on the energetics and electronic structures of BC$_2$N nanoribbons is desired.

We also performed the first principles calculations based on the density functional theories within the projector augmented wave (PAW) method  \cite{Blochl1994} and LDA \cite{Perdew1981} implemented in {\scriptsize VASP} code.\cite{Kresse1996prb,Kresse1996cms}
The BC$_2$N nanoribbons shown in Fig.\ \ref{fg:RibbonStructure} have the flat bands BC$_2$N nanoribbons when the outermost atoms are terminated by single H atoms.
According to the results within the first-principles calculations, the on-site energy, $E_{\rm B}$, seems to be larger than the hopping integral, $t$.
However, these issues are beyond the scope of the present paper and will be reported in future publications.

\section{Summary}

We theoretically studied the electronic and magnetic properties of BC$_2$N nanoribbons with zigzag edges using the tight binding model and the Hubbard model.
We showed that zigzag BC$_2$N nanoribbons have the flat bands and edge states when the atoms are arranged as B-C-N-C along the zigzag lines.
The length of the flat bands depends on the atomic arrangement, which can be explained by the shift of the Dirac point in BC$_2$N sheet from the K point of the honeycomb lattice.
The charge distribution also depends on the atomic arrangement.
When the outermost sites are occupied by C atoms, the charge distributions of the edge states in zigzag BC$_2$N nanoribbons are resemble to that of conventional edge states.
When the outermost sites are occupied by B and N atoms, on the other hand, the charge of the edge states distribute over both sublattice sites, which are different from those in conventional graphene zigzag edge.
We also showed that such charge distribution causes different magnetic structures.
When the outermost sites are occupied by C atoms, we observed the ferrimagnetic order at the edges independent of the magnitude of the on-site Coulomb interaction, which are resemble to that of conventional edge states at the zigzag graphene nanoribbons.
At the zigzag edge where the outermost sites are occupied with B and N atoms,  on the other hand,
the ferromagnetic structure appears when the site energies are larger than the on-site Coulomb interaction and the magnetic structures turn into the ferrimagnetic with increasing  the on-site Coulomb interaction.

\section*{Acknowledgments}

The authors acknowledge H.\ Imamura, Y.\ Shimoi, H. Arai, H. Tsukahara, K.\ Wakabayashi and S.\ Dutta for valuable discussions.
This research was supported by the International Joint Work Program of Daeduck Innopolis under the Ministry of Knowledge Economy (MKE) of the Korean Government.

\appendix

\begin{figure}[t!]
\centering
\includegraphics[width=7cm]{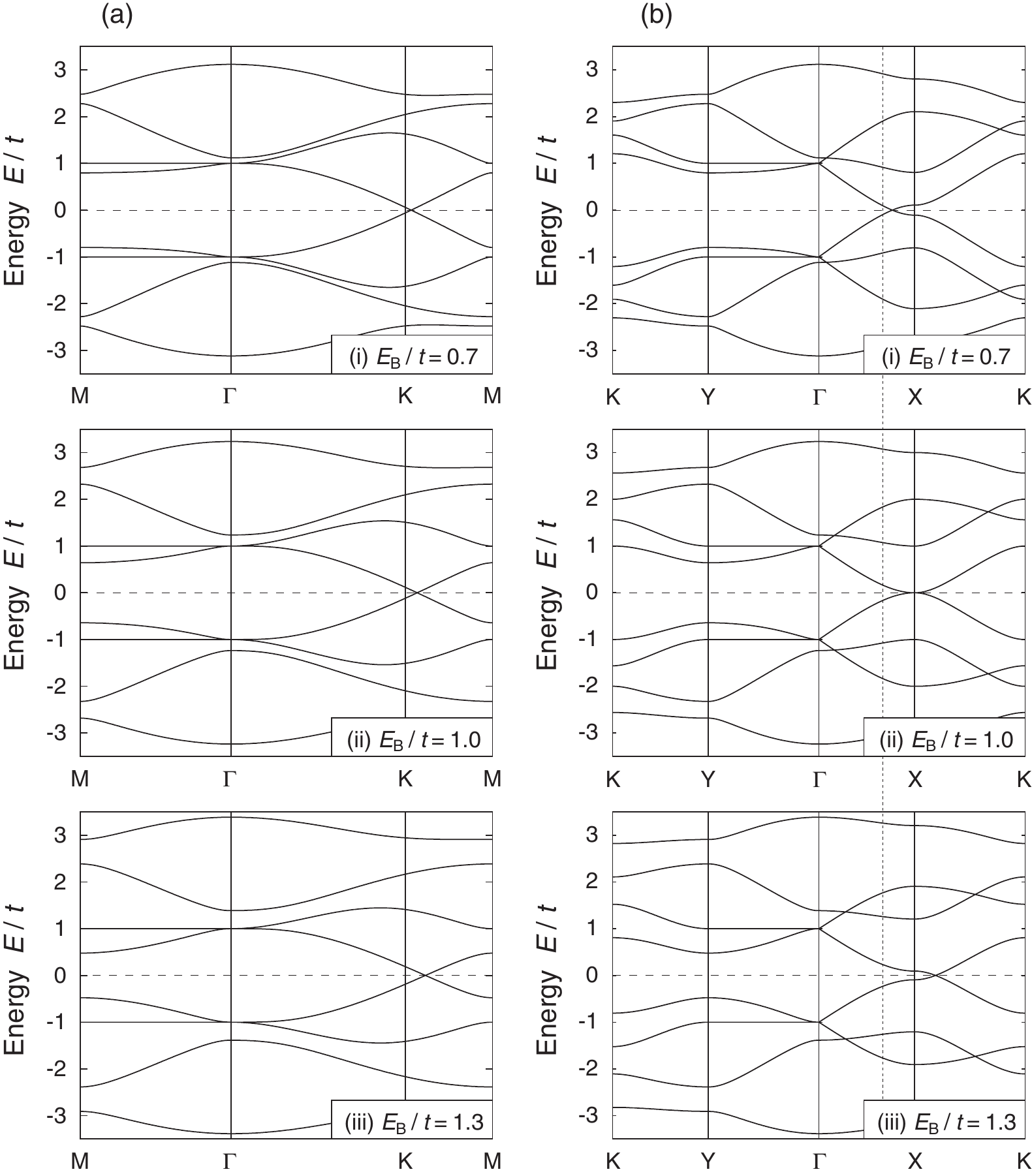}
\caption{
(a) The band structures of BC$_2$N sheet which consists the model-7 and 8 BC$_2$N nanoribbons. 
(b) The band structures of BC$_2$N sheet which consists the model-9 and 10 BC$_2$N nanoribbons. 
The vertical dotted line represents the position of the K point of the honeycomb lattice.
}
\label{fg:add-2d-band-2}
\end{figure}

\section{BC$_2$N nanoribbons with B-B and N-N bondings}

In this appendix, we shall present the results of the BC$_2$N nanoribbons with B-B and N-N bondings.
In the left sides of Figs.\ \ref{fg:add-Structure-LDOS} (a-d), the schematic illustrations of BC$_2$N nanoribbons with B-B and N-N bondings are presented.
The calculated band structures for $N=10$ and 30 are summarized in Figs.\ \ref{fg:add-N10-band} and \ref{fg:add-N30-band}, respectively.
We found that the length of the flat bands in the model-7 and -8 nanoribbons are shorter than that of zigzag graphene nanoribbons, while the flat bands in the model-9 and 10 nanoribbons are longer than that of zigzag graphene nanoribbons.
It should be noted that the flat bands of the model-9 and 10 nanoribbons for $E_{\rm B}/t\geq 1.0$ cover the whole Brillouin zone.
Calculated LDOS at $E=0$ are given in right sides of Figs.\ \ref{fg:add-Structure-LDOS} (a-d) with the magnitude of LDOS are represented by the area of the circles.
The flat bands correspond to the edge states.
We found that the electric charge distributes with similar manner to those shown in Fig.\ \ref{fg:LDOS}.

As we discussed above, we shall consider the band structures of BC$_2$N sheets.
Figure \ref{fg:add-2d-band-2} (a) show calculated band structures of BC$_2$N sheets  which consist the model-7 and 8 BC$_2$N nanoribbons.
Corresponding Brillouin zone is shown in Fig.\ \ref{fg:Hex2D-band} (c).
With increasing  $E_{\rm B}$, the Dirac point deviates from K point toward the M point.
Therefore, the Dirac point in the one dimensional Brillouin zone shifts toward the $\Gamma$ point, resulting in shorter flat bands.

Figure \ref{fg:add-2d-band-2} (b) show calculated band structures of BC$_2$N sheets  which consist of model-9 and 10 BC$_2$N nanoribbons.
Corresponding Brillouin zone is shown in \ref{fg:Rec2D-band} (c).
In this figure, the vertical dotted line represents the position of the K point of the honeycomb lattice.
In this case, the Dirac point is deviated from the K point of honeycomb lattice toward X point for $E_{\rm B}/t=0.7$.
Therefore, the length of flat band becomes longer than that of zigzag graphene nanoribbons.
For $E_{\rm B}/t=1.0$, the dispersion becomes massive at the X point.
Then, the flat bands cover the whole Brillouin zone.
With increasing  $E_{\rm B}$, we observed the Dirac cone but the Dirac point are located on the X-K line.
In the one dimensional Brillouin zone, there are states with $E=0$ at $k=\pi$ and the flat bands remain covering the whole Brillouin zone.
Therefore, the length of flat bands depend on the atomic arrangement as discussed before and the change in the length of flat bands can be understood as the shift of the Dirac point in BC$_2$N sheet from the K point of honeycomb lattice.

\end{document}